\documentclass{article}
\usepackage{amsmath}
\usepackage{amsfonts}
\usepackage{graphicx}
\begin{document}
\title{Science and the Future: Introduction}
%
%

\author{Angelo Tartaglia \\
Department of Applied Science and Technology, Politecnico di Torino, \\
Corso duca degli Abruzzi 24, 10129, Torino, Italy \\
\and INFN, Via Pietro Giuria 1, 10126, Torino, Italy}

\maketitle

\begin{abstract}
  The contradiction between physical and economical sciences concerning the
growth of the production/consumption mechanism is analyzed. It is then shown
that if one wishes to keep the security level stable or to enhance it in a
growing economy the cost of security grows faster than the gross wealth. The
result is a typical evolution in which the net wealth increases up to a
maximum, then abruptly collapses. Besides this, any system of relations
based on a growing volume of exchanges is bound to go progressively out of
control.

The voluntary blindness of the ruling classes toward these facts is leading
our societies to a disaster. This fate is not inescapable provided we learn
to dismantle the myth of perpetual growth.
\end{abstract}
\section{Foreword}
\label{intro}
The history of the human kind on the planet Earth is extremely short as
compared to geological or cosmic times, but is felt as being already quite
long in the human perception. In fact the part of our common past properly
considered as historical has had a duration of approximately 400
generations: nothing, in a world where every year some 125 million new human
beings are born (more than 343,000 per day). Even in such a short time span
the impact of humanity on Earth, never irrelevant (as some of the talks in
these conference will show), has undergone a sharp raise starting from some
fifteen generations ago. The birth and development of modern experimental
science (mainly physics in all specifications) spurred technology, in a
previously unthinkable way. These new powerful tools have been made
available to a species defining itself as "intelligent"; a species whose
behaviour has at all times been conformed to non-scientific criteria and
directed to not properly rational ends. The \textit{de facto} behaviour of
the human beings has however ever since been covered, like by a coat or a
sort of Grecian theater mask, by a conceptual framework intended to justify
it or even motivate it \textit{ex post}. Any human society, since Sumerian
times up to the present day, has worked out an ideological
self-representation that "explains" it is right that it happens precisely
what happens.

\begin{figure}
\centering
\includegraphics[width=12cm]{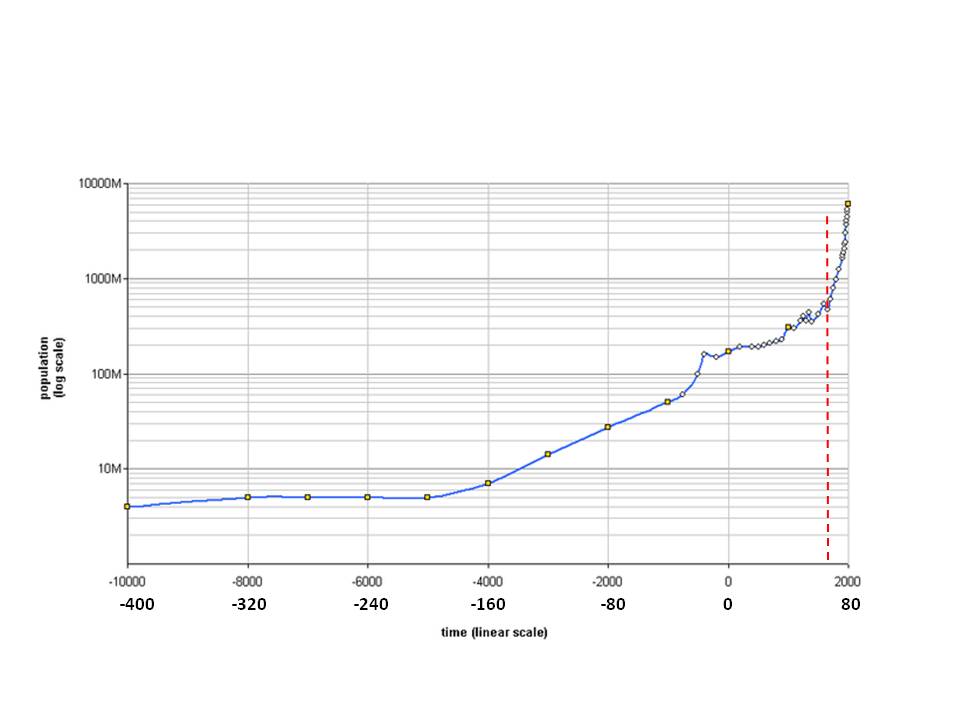}
\caption{Growth of the world population in time; the lower line in the abscissae measures time in conventional human generations (25 years). The dashed line marks the date (some 14 generations ago) from which the modern "explosion" of the population starts, accompanying the industrial revolution.}
\label{fig-0}       
\end{figure}

\section{The role of science and the problem of limits}
\label{sec-1}
The development of the scientific knowledge and method has progressively
brought to evidence some inconsistencies or plain contradictions between
ideological visions and the material advancement of human societies. These
inconsistencies and contradictions have in general to do with the concept of
limit, meant as material constraint. In ideological views, different and
changing in time as they are, the very idea of "limit" has never been
popular: one meets sheer refusal (we shall see an interesting example
tomorrow evening in the "Last call" movie) or, at times, the recognition
that 'well, material constraints do exist, but they are exceedingly far from
our present condition, so that they are and must be uninfluential on the
decisions we are called to assume here and now'.

Another way to describe this attitude could be that mankind tends and
pretends to live exclusively in the present, removing any concern about a
non-immediate future. In short, a time dimension with a thickness close to
zero. This behaviour is what we currently expect from non-human animal
species; the point however is that the human being, especially after the
onset of modern science, has got tools and methods enabling to look, in a
reasonably reliable way, even into a non-immediate future. What can be seen
in that, scientifically fathomable, future are developments contradicting
the perspectives advertised by ideology (here I am not referring to any
peculiar political vision, rather to the most common way of thinking of
humans and in particular of the members of the "ruling class").

The importance and inescapability of material constraints is indeed rather
obvious and has been perceived as such all along the history of human
thought. That awareness, however, has in general not exerted any real and
effective influence on actions. After the start of the industrial revolution
we find an example of such awareness in the positions of rev. Malthus, who
apparently concentrated on the demographic issue only. Only in very recent
times, however, "exact" sciences have tackled the problem of constraints:
general public was involved in 1972, when \textit{The limits to growth }was
published. We shall recall that event tomorrow afternoon. The report,
produced by a team of the MIT (Donella Meadows, Dennis Meadows, J\o rgen
Randers, William Behrens III) and commissioned by the Club of Rome, was
promoted by Aurelio Peccei, a native of the city where we are now.

\textit{The Limits to Growth }was immediately attacked (ideologically) both
from the "left" and from the "right": the idea that material growth of the
economy could have physical limits was \textit{unacceptable}. This pretended
"unacceptability" is related to a sort of human grandiose delusion that
should be matter for psychologists, but is in fact deeply rooted in a
culture, especially European, now spread worldwide by right of conquest,
which maintains that matter is subordinate to "spirit". "Spirit" here is not
the one of religion, rather, if anything, the one of Hegelianism; but I
neither can nor want to enter philosophical considerations, so let me drop
this issue.

It's a given that matter: a) exists; b) evolves according to laws which
depend neither on philosophical nor on economical schools and are not at all
influenced by political affairs. Be they acceptable or not, no parliament
can reform the physical laws and they are not even listed in the stock
market.

Since 1972 to the present day a true scientific literature has been produced
concerning the material constraints human societies cannot avoid to come to
terms with, even though they are not willing to do so. A trivial search on
an academic search engine pinpoints in a split second more than 2,000,000
scientific papers on the climate change issue, produced in universities and
research institutions. If you try a query on the \textit{peak oil} you find
not less than 1,000,000 articles, and so on. For comparison, a subject like
\textit{superconductivity }gives less than 650,000 answers and a general
term, typical of my normal research activity, like \textit{cosmology }%
produces less than 500,000 titles.

National and international conferences have grown in number and frequency
and appeals have been proposed to reroute the world economy and to curb the
mythical "growth". A prestigious example has been the Stockholm memorandum,
signed in 2011 by 18 Nobel laureates. Positive intentions have been
expressed hither and thither; some states have signed agreements such as the
Kyoto protocol of 1997, or have defined targets to be pursued, like the
20-20-20 directive of the European Union; and so on. Something has started
to move and is moving. In substance, however, the problem has not been
tackled in its root and the global situation has worsened; in the best case
one worries about the symptoms, not about the disease.
\section{Economics}
\label{sec-2}
What I referred to above, when writing of the boom of specialized
literature, were the sciences of nature, all more or less "exact", provided
this attribute makes sense. Those sciences deal with material quantities and
measure in units like kilograms and Joules (or kWh). In that context not
anything concerning the future is clear and unambiguous; there are uncertain
and even controversial issues. The bases, however, and the rules of the game
are definitely certain: the discussion may be about technical problems and
ultimately on details, even if non-marginal ones (we shall see in the next
days of the conference), but the substance is unique and clear.

But not only the sciences of nature exist; other disciplines can be
qualified as sciences too, because of the method and accuracy in analysis
and application. In particular, in the matter that we shall be discussing in
these days a fundamental role is played by economics. A layman, as I am on
this respect, is led to say that the object of economy as a science is human
behaviour, like in a sort of social psychology applied to exchange
relationships of goods and services within a given society; I should include
also the production of goods and services directed, in a way or another, to
the exchange. If it is really so (I beg the pardon of economists) analyses,
"laws", mathematical formalization are necessarily statistical and the
observations are mainly empirical, axioms are quite different from those of
physics or even more of mathematics. In economics basic hypotheses, more
than axioms, betray also the (human) world views typical of one or another
society and lack the logical cogency of the axioms of natural sciences.

Having said that, economics too has enormously grown in complexity, internal
differentiation and sophistication of the technical tools it uses. Economics
is formulated in refined mathematical terms, manipulates quantitative data
and produces numerical predictions. Of course a person like me, with a
physicist's (and engineer's) mentality, gets puzzled looking at the numbers
appearing in newspapers, even in the pages dedicated to economy and finance;
not because of the values in themselves, but rather because of the
systematic absence of anything like an \textit{uncertainty }interval \textit{%
\ }or, when modelling is used (i.e. almost always), of a \textit{confidence }%
interval for predictions: we shall discuss this aspect on Thursday. My
perplexity, as I have said, concerns communication through the media, but
that is the channel which conveys messages to the general public.

Of course economy deals with material quantities, the ones measured in kg,
kWh and so on, but the most important quantity is in the end \textit{money};
at least, everything ends being expressed in \textit{monetary }terms. Now,
again, a physicist cannot help but remark that money is fundamentally a
\textit{conventional }quantity\textit{, }that expresses, within a given
society, the distribution of the buying power (to a great extent coinciding
with power, without further specifications), i.e. of the right of access to
goods and services globally produced. In other words, there is an objective
poverty, consisting in the scarcity or material inaccessibility of some
good; then there is a "social" poverty depending on the unequal distribution
of the "access titles". In our globalized world the difference between the
two forms is more and more evanescent. In conclusion, though it appears to
be paradoxical, the poor is poor because "he agrees to be poor" even though
he does not know: he partakes in a giant Monopoly game whose rules he, more
or less consciously, accepts. All this has nothing to do with the laws of
thermodynamics, with the conservation of matter and energy, with
climatology, with the laws of electromagnetism, and so on. Disparities and
the fair or unjust distribution of the "access titles" is a match internal
to mankind, but that match is played on a material stage; the rules of the
game are, with more or less difficulty, pliable, the playing ground and its
constraints not at all.

Economical science, by its nature, has to deal with both the material
context and the internal affairs of human societies. It seems however, at
least to an inexperienced person as I am, that the attention is most often
focussed on human dynamics rather than on the material background. I shall
fetch an image from a field very distant from economy, but much more
familiar to me. Einstein succeeded in describing the gravitational
interaction by a very famous (among specialists, of course) equation of
which he used to say that the two sides (linked by the = sign) were quite
different: one was "marble", and it was the one rigorously and exclusively
cast in mathematical and in particular in geometrical terms; the other was
"wooden", and it was the one containing matter and energy, both entities
expressed in mathematical forms without being mathematical in nature. In an
entirely different area and by pure analogy I think I can say the proper
domain of economy too has two sides: one "marble", which is the side of
material bases; the other "wooden", which is the side of social dynamics. My
feeling (and may be this conference will help to discuss the issue and make
me change my mind), is that very often economy mistakes "marble" for "wood",
without, by that, "marble" ceasing to be what it is.
\section{The problem of growth}
\label{sec-3}
On the background described so far we find the prickly problem of "growth".
"Growth", at first sight, is the most often used word on the press, by
politicians, by economists when the discussion is about the world "crisis",
started in 2008 and not yet ended, and about the way out from it. On the
nature of that "crisis" I think this conference will express peculiar
viewpoints; for the time being let us limit ourselves to the question:
growth of what?

In the debate among social parts and in the journalistic and popular
language the growth people want to start up again is of consumption, of jobs
(of consumption in order to make jobs grow), of investments, of production,
of the "confidence" of "markets" and so on. When we come to the numbers, all
that is recapped and summarized in one single parameter: the Gross Domestic
Product, in short GDP. The latter, in the popularization of the media, is
not even expressed by an absolute value, but only as an annual percent
growth (lately even decrease): "GDP is at x\%". Literally speaking this is
nonsense; for the majority it means that things are good or bad, without
worrying about why they are good or bad. Long since there is an active
debate on the reliability of the Gross Domestic Product as an  indicator of
the socioeconomic wellness and prof. Giovannini could tell us much about the
subject; as a matter of fact people goes on using GDP.

As I understand it, and I am a layman, GDP measures the volume of material
and immaterial exchanges within a given economic system, provided they have
a monetary value. That is the quantity whose growth everybody invokes.
Stated in these terms one may have the impression that the issue is
essentially an internal affair of human societies, all the more so, since,
as I have said, the exchange can also be immaterial: a violin lesson
contributes to the formation of the GDP, provided it is paid for (what is
free of charge, in this context, besides not been counted, is at the least
an extravagance).

Beyond the political and journalistic language, I can guess that the
theoretical framework that includes the growth, rests on two postulates.

\begin{itemize}
\item a) The vitality of any human society can last in time if there are
disparities among its members: in an electric circuit no current circulates
if the potential is the same everywhere.

\item b) Economic disparities, however, have drawbacks, dangerous for social
stability. If you want to find a solution to these drawbacks or at least
keep their effects under control, the volume of the economic exchanges
\textit{must }keep on growing: some would say that in order to maintain
social differences, without producing lacerations, the global "wealth" is
bound to a perpetual growth.
\end{itemize}

Is this the essence of an amateur theory of growth? Maybe.

However the problem is that any "exchange", including the apparently
immaterial ones, has unavoidably a material base. By consequence the growth
of the Gross Domestic Product is necessarily and in any case a growth of the
amount of matter manipulated and transformed every year and an increase in
the number of transformations to which even a fixed amount of matter is
subject every year (at least part of the present audience has probably
caught here a reference to the growth of the energy demand). When we speak
of matter and its transformations we face rules of the game on which we have
no control; they are, as I have already reminded, totally insensitive to
political and even social affairs, to the spread, to the stock market
quotations, as well as to the unemployment rate. At this stage the
contradiction becomes manifest and, it is to be said, on one side we may
heap opinions, on the other we find very stubborn \textit{facts}.

There are three ways to manage these contradictions by the supporters of
growth:

\begin{itemize}
\item reject the problem;

\item recognize that the contradiction exists, but maintain that its
practical effects are delayed to back of beyond future;

\item propose a socioeconomic system where the material base remains
constant and the GDP\ increases in its immaterial part only.
\end{itemize}

No doubt that the first "solution" is the most widespread. It includes those
people who look at the issue of growth as at a set of "environmental
problems"; for these people the "environment" becomes rather a means to
promote the "growth" of the economy: "let us invest in 'green technologies'
and GDP will start to grow again". It has barely to be said that the
"environment" has no problem at all; rather it is \textit{we }only who have
problems. This attitude has the rationality of a guy with a broken leg who
plans to participate to a high jump competition.

The second option implies an always present debate on the closeness or not
of the walls that prevent an indefinite "growth". We are going to discuss
this matter tomorrow, but I think I can anticipate that by now the limits
are indeed very close, even on the time scale typical of human affairs. In
any case recognizing the problem, then differing it, for sure does not solve
it. It recalls a bit the sentence "\textit{In the long run we are all dead}"
by John Maynard Keynes, or the "\textit{apr\`{e}s nous le d\'{e}luge" }by
Louis XV of France, or even the "\textit{seize the day}", modern version of
Horace's \textit{carpe diem}. Rationality dwells elsewhere.

\begin{figure}
\centering
\includegraphics[width=8cm]{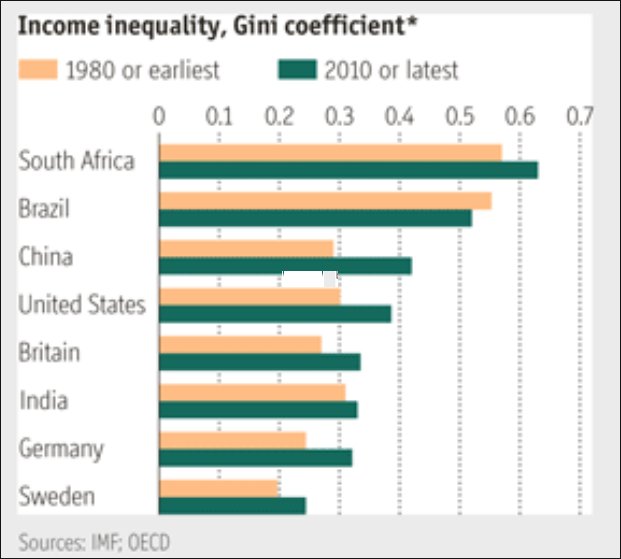}
\caption{Time evolution of the Gini coefficient in the world. The Gini coefficient measures the inequality of the income distribution. The situation has worsened almost everywhere in 30 years; Brasil, where the inequality is very high, is an exception.}
\label{fig-g}       
\end{figure}

Coming to the third solution some might propose, maybe we will discuss it in
the next days. There are however various inconveniences. In order to allow
GDP to grow keeping the amount of manipulated matter fixed, one must let the
number of "manipulations" performed on a given amount of matter increase,
but also the "manipulations" unavoidably have a material (in particular
energy) content which is subject to the same constraints as the whole
material world. Furthermore there is a more subtle limit: if we want the
volume of exchanges to grow keeping the exchanged quantities fixed, we must
increase the exchange \textit{velocity}, but velocity too has an
insurmountable upper limit in any given physical system. I'll come again on
this later. One might think to only build on virtual exchanges, like the
more and more happens in the world of finance, but even so a remark is
inescapable: if quantities remain stable but the swirling exchange of
"access titles" grows, what grows in the end, like in a worldwide Monopoly
game, are inequalities. There is evidence of this phenomenon in our world,
even though, at the moment one pretends quantities also to go on growing:
there is a limited number of subjects, in prospect the more and more
limited, who own disproportionate "withdrawal rights" against a growing
number of subjects who are losing their economic autonomy; see figure (\ref{fig-g}). Of course this is
the type of growth that produces more immediately explosive effects.
\section{An example: governing an ever-growing system}
\label{sec-4}
Mine is an introductory relation, so that I\ limit this talk to a review of
various problems and topics without examining them in depth. Thorough
analyzes will be made in the course of the conference. Nonetheless, before
closing, I would like to develop an exercise, touching upon features of the
"growth" mechanism, which are usually not considered or remain in the shade.

Everybody knows that there is a problem with, and a live debate on,
anthropogenic climate changes; that there is a problem with the depletion of
energy resources and with the scarcity of mineral resources; that there is a
problem with fresh and clean water; degradation and reduction of available
farming soils is an issue often pointed out to the public, and so on. About
these topics we shall discuss in the next days.

Now I would like to draw your attention on the control and safety of a
growing complex system. I will do it in a simplified way trying to reduce as
much as possible the technicalities and to evidence the fundamental
machinery.
\subsection{A network of exchange links}
\label{sec-41}
An economy, as we have seen, is based on a system of exchange relations. The
growth that many invoke implies also an increase of the exchange
opportunities, the number of companies, the market segments. We may think of
human beings exchanging something among them; or villages connected by
pathways; or factories and markets connected by roads and information
channels.... We may sketch all this thinking of a three-dimensional network
(I should say a \textit{graph}) whose vertices (or knots) are the things I
have just mentioned. Let us suppose that the vertices are in the number of $N
$. The produced wealth is not \textit{per se} depending on the number of
vertices, but rather on the exchange fluxes among them and, in an abstract
market, all knots should be on the same footing. In other words the best
conditions for such economy would be obtained if any single knot could
exchange something with any other single knot in the network. The relevant
quantity emerging from these considerations is the number of \textit{links}
among the knots. Now, the total number $\mathfrak{N}$ of possible links
among $N$ vertices is given by a simple formula of the binomial development:

\begin{equation}
\mathfrak{N=}\frac{N\left( N-1\right) }{2}  \label{relazioni}
\end{equation}

You may notice that when $N$ grows, $\mathfrak{N}$ grows faster and
precisely with a quadratic law: to 2 vertices it corresponds 1 possible
link; 3 vertices, 3 links; 4 vertices, 6 links; 5 vertices, 10 links; and so
on.

So far I have spoken of \textit{possible} links, but wealth originates from
\textit{actual} exchanges, not from \textit{possible }exchanges. Once a
channel is available, the flux through it will initially be small, but, if
wealth, whatever it is, depends on the total flux and people wants it to
grow, then also the flux on each single link must grow. If we call $n_{i}$
the number of transactions or exchanges or trips per year through the $i$-th
available connection, the total annual volume of exchanges $\Phi $ will of
course be

\begin{equation*}
\Phi =\overset{\mathfrak{N}}{\sum\limits_{i=1}}n_{i}
\end{equation*}%
The formula may be simplified a bit using the average annual flux through a
link, $<n>$, and becomes

\begin{equation}
\Phi =<n>\mathfrak{N}  \label{PIL}
\end{equation}

If I wouldn't be afraid of the economists' reproaches I would be tempted of
saying that $\Phi $ is proportional to the Gross Domestic Product of our
system; anyway I shall say that it is proportional to the wealth $R$
produced by the system:

\begin{equation}
R=\rho \Phi  \label{ricchezza}
\end{equation}%
The $\rho $ factor is a parameter carrying all the ambiguity of quantities
measured by money, however if $R$ is\textit{\ bound }to grow, as it is
maintained by many, the same will do $\Phi $ and this will be possible
either through any of the two factors in formula (\ref{PIL}) or through both
jointly. In any case the growth of the number of possible links, $\mathfrak{N%
}$, requires the growth of the number of "poles" $N$.

Coming to the average annual flux through a link, in its evaluation also
those links enter which are in principle accessible but not operating and
along which the flux is zero; this fact implies that $<n>$ can grow: a)
letting the number of active exchange links increase up to the greatest
possible value for a given number of vertices [formula (\ref{relazioni})];
b) letting the flux through each single channel grow. Now, option a) has to
comply with the maximal capacity of a single vertex, which in turn depends
on the nature of the vertex: if we have to do with an electronic
communication network the knots are electronic devices (servers, routers,
laptops...) each of which has a maximal capability of managing separate
connections, because of the physical constraints typical of the employed
materials, of the commutation speed, etc. In order to avoid a typical
objection of those who deny the existence of limits, I immediately specify
that I am not referring to specific machines of the present technological
generation that would then be surpassed by the next generation, then by
further one, then.... . I am referring to ideal machines working only on the
base of the physical principles which rule the atomic physics (the branch of
physics at the base of the applications of electronics) and invariably imply
the saturation of the capacity of \textit{any} calculation or data
processing device.

If the vertices of the graph are human beings the problems are the same but
for the fact that saturation is reached much earlier.

Let's come to option b). Growth requires an increase of the flux transiting
over every single link of the network. Here I will not spend too many words:
be it a road or an optical fiber, a railway or a satellite communication
channel, under a continuous push towards growth the trend in time of the
flux is represented by a logistic curve, like the one shown in figure (\ref%
{fig-1})
\begin{figure}
\centering
\includegraphics[width=12cm]{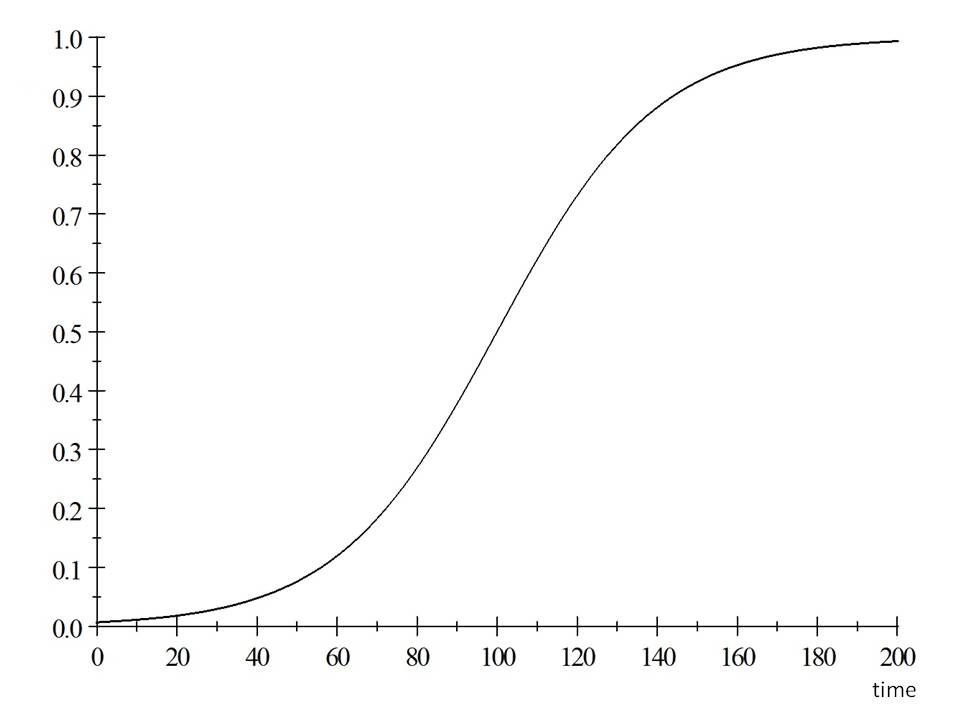}
\caption{A typical logistic curve
describing the saturation in time of a channel in which a continuous thrust
towards the increase of the flux is present.}
\label{fig-1}       
\end{figure}

The saturation of all or part of the links is unavoidable and the saturation
level depends on the nature of the link. Under a continuous pressure towards
growth the links tend to saturate and consequently push towards the
saturation of the number of links manageable by every single vertex. Finally
the growth must necessarily end being transferred on the global size of the
network, i.e. to the number of vertices. Applying all this to the whole
human kind it turns out that the "growth" must in the end be demographic
also; if the number of "vertices" of the graph does not grow and "wealth"
\textit{must }grow, the system saturates and jams. On the other hand the
growth of the number of vertices has to comply with the physical constraints
of the environment where the growth takes place. The debate could go on and
evolve for long, but would not be \textit{per se} new: after all I have just
presented in a more abstract way the subject of "\textit{The limits to growth%
}" and in the next days various specific lectures will do better than I do.
However the aspect I want to highlight in the behaviour of our network is
not the one of the physical constraints.

\subsection{The problem of safety/reliability}
\label{sec-42}

Let us come again to the number of exchanges that happen per year on the
average through a link of our network, $<n>$; each exchange, if it is real,
is unavoidably associated with a probability, $\varepsilon $, of undergoing
some inconvenience, breakdown, accident, malfunction. Over big numbers this
probability becomes statistically a global real annual number of
"inconveniences", $\nu $, given by

\begin{equation}
\nu =<\varepsilon ><n>\mathfrak{N}=<\varepsilon >\Phi  \label{rischio}
\end{equation}

If we have considered $\Phi $ as being proportional to the gross "wealth"
produced during the year, we may think $\nu $ being proportional to the
damage $D$ overall suffered because of the happened inconveniences:

\begin{equation}
D=\rho \nu  \label{danno}
\end{equation}%
Strictly speaking damage is not simply proportional to the gross wealth: in
general $<\varepsilon >$ (which is contained in $\nu $) depends in turn on $%
\Phi $. Let us think for instance of a road with vehicles travelling on it;
keeping technology, road and driver fixed, the probability of individual
accident grows with the flux of vehicles.

Summing up, our system behaves more or less like a ferromagnet:
magnetization grows more than proportionally with the intensity of the
magnetizing field and tends sooner or later to reach a fixed saturation
value. The dependence of $\varepsilon $ on $\Phi $ changes according to the
type of "inconvenience" and sometimes it may also be absent: let us merge
all this into a hypothetical average of all averages in order to arrive to
the $<\varepsilon >$ of (\ref{rischio}), that will now represent the total
"risk". In general, and when we are far from saturation, we expect
insecurity to grow while wealth grows, i.e. we expect the ratio between risk
and gross benefit to worsen. Of course this is an unwanted and not at all
relished effect; is it possible to nullify it? Be it car crashes, engine
breakdowns, noise injected into a communication channel, robberies, frauds,
bribery or else, one may hope to keep everything under control below a
previously fixed threshold, by means of an improvement of the efficiency of
the system, i.e. of the technologies, the methods and the controls. And here
the problem of the \textit{cost} of all that comes to light.

The efficiency of a system, interpreted as the ratio between desired effects
and globally produced effects, cannot grow indefinitely. It stands to reason
that the maximal abstractly attainable value would be $1$; actually even the
maximal value attainable in principle is less than $1$: "side unwanted
effects" are \textit{necessary}, whatever the quality of the employed
technology is. Engineers and physicists in the audience have undoubtedly
realized that I am speaking of the first and second principles of
thermodynamics: that kind of laws that parliaments do not vote and cannot
amend, on which autocrats have no power, which are totally unaffected by the
subtleties of rhetoricians.

\begin{figure}
\centering
\includegraphics[width=12cm]{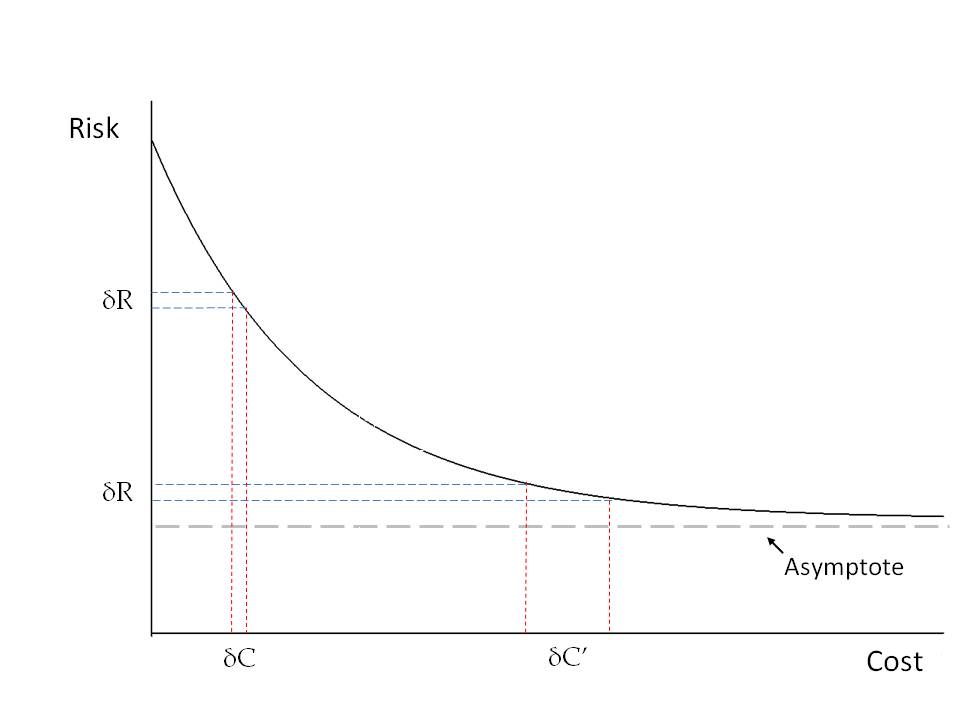}
\caption{Asymptotic decrease of "risk"
with increasing cost. The same $\delta R$ corresponds to different $\delta C$}.
\label{fig-2}       
\end{figure}

Consistently with the example I am trying to work out, rather than to
efficiency I will refer to the probability of "inconvenience" for each
single exchange: let us say that we will keep an eye on \textit{risk}. Since
$<\varepsilon >$ grows more than proportionally with the flux $\Phi $ in the
network, we shall try to compensate the unbalance reducing the risk for a
given flux. This aim, as I have already said, can be pursued improving
"efficiency" by various technological, behavioural, regulatory measures. All
this has a cost, no matter how you evaluate it, and the typical trend of
risk as a function of "cost" is schematically shown in figure (\ref{fig-2}).

What happens is well known: at the beginning, when the efficiency is low and
the risk is high, one obtains relevant results with limited cost, but once
the risk is already relatively low further improvements have bigger and
bigger costs. Let us swap the axes in the figure (or rotate it by $90%
{{}^\circ}%
$ on the left): one clearly sees that when the "risk" tends to the asymptote
the cost diverges. Of course the cost that we can call 'of security' must be
subtracted from the gross wealth $R$, proportional to $\Phi $, so that the
actually available wealth, $\mathcal{U}$, decreases.

So far the colloquial and qualitative description of what happens. Let us
now try a numerical exercise. Start from the gross wealth $R$. According to
the wishes and also to the "doctrine" $R$ (if you prefer: the GDP) should
steadily grow with a geometric progression; in practice its evolution in
time should be described by an exponential function:

\begin{equation}
R=R_{0}e^{\chi t}  \label{expR}
\end{equation}

We should choose a value for $\chi $. Initially I thought of $2\%$ per year.
However, it occurred to me of reading, on the newspaper La Stampa of
September 5 this year, an article by a professor of economics of the Turin
University, Prof. Mario Deaglio, inspired by some sentences of a singer
(Jovanotti) about the "wonderful" (environmentally friendly) growth. Prof.
Deaglio writes\footnote{%
The English translation is by the author of the present Introduction.} "a 1
per cent growth per year has nothing 'wonderful', a 5 per cent growth would be unattainable. Jovanotti's 
'wonderful growth' corresponds to an average annual increase
of the gross domestic product (the notorious GDP) in the order of a long
term 2.5-3 per cent. Such growth rate has to be steady, non-dehumanizing,
not particularly consumeristic." That growth rate, according to Deaglio,
would include both the productivity and the employment increase; no material
quantity is mentioned: no tons, no kilowatt hour, no mundane variable that
could disturb the "wonderful growth". In line with this approach, I will
then choose

\begin{equation}
\chi =0.03/year  \label{crescita}
\end{equation}%
which of course corresponds to a similar growth rate for the exchange volume
$\Phi $.

We must now deal with (\ref{rischio}) and (\ref{danno}). The majority will
expect the total risk to remain constant or even to decrease while the
produced wealth increases; hence we deduce that people will act upon what
I'll generically call the \textit{technology }in order to try and obtain a
\textit{decrease} of $<\varepsilon >$ in time big enough to compensate for
the growth of $\Phi $, i.e. people will require that

\begin{equation}
<\varepsilon >=\varepsilon _{0}e^{-\chi t}  \label{probtempo}
\end{equation}

Strictly speaking this result is impossible: no real transfer may happen at
\textit{zero} risk. Let us however neglect this remark and let us pass to
consider the total cost of security $\mathcal{C}$, that is the quantity
schematically represented on the abscissae of fig.~(\ref{fig-2}). The trend
depicted there may be expressed by the formula:

\begin{equation}
\mathcal{C=}\frac{\kappa }{\varepsilon -\varepsilon _{\ast }}\Phi =\frac{%
\kappa }{<\varepsilon >-\varepsilon _{\ast }}\frac{R}{\rho }  \label{costo}
\end{equation}%
or even by a number of more complicated expressions. This is just an
exercise, so let us stay happy with (\ref{costo}) where one sees both the
proportionality to the total exchange flux, then in practice to the produced
wealth, and the divergence in correspondence of the asymptotic value $%
\varepsilon _{\ast }$. Let us imagine to start from a situation at time $0$
where the cost of safety is a fraction $\psi $ of the total wealth, so that
we may write:

\begin{equation}
\mathcal{C=}\psi \frac{\varepsilon _{0}-\varepsilon _{\ast }}{<\varepsilon
>-\varepsilon _{\ast }}R  \label{costoR}
\end{equation}

Making the time dependences explicit, the formula becomes

\begin{equation}
\mathcal{C=}\psi \frac{\varepsilon _{0}-\varepsilon _{\ast }}{\varepsilon
_{0}e^{-\chi t}-\varepsilon _{\ast }}R_{0}e^{\chi t}  \label{espilicito}
\end{equation}

Let us try and see what happens if the initial cost is $1\%$ of the initial
wealth and the initial risk is $100$ times bigger than the asymptotic value.
Figure (\ref{fig-3}) shows the time evolution both of $R/R_{0}$ and of $%
\mathcal{C}/R_{0}$:

\begin{figure}
\centering
\includegraphics[width=12cm]{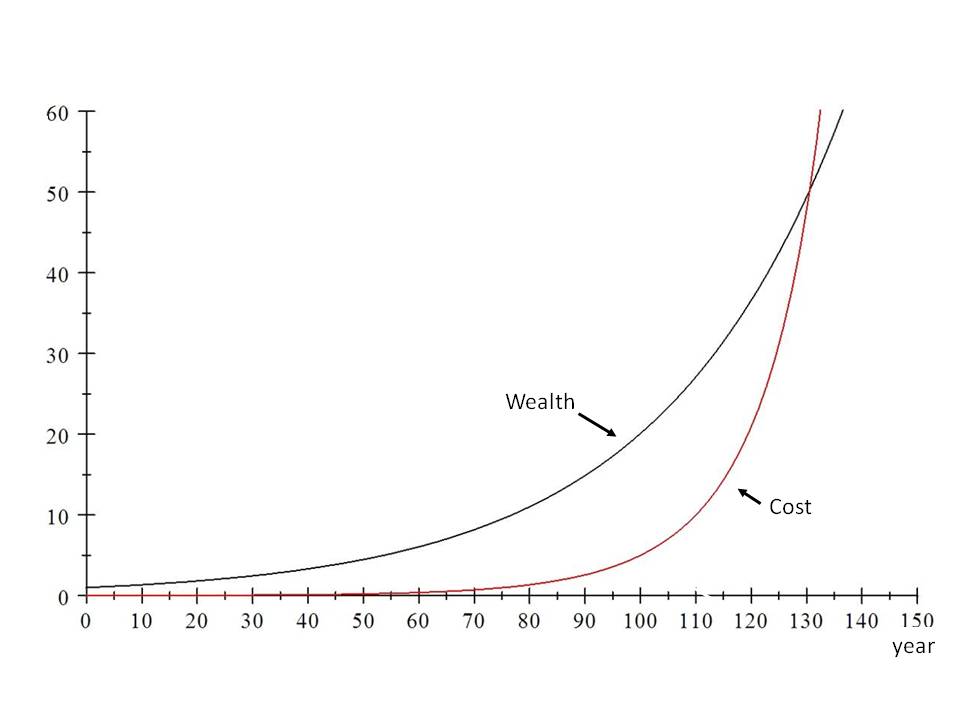}
\caption{The black curve represents the
growth of the gross wealth at a rate of 3\% per year. The red curve is the
time trend of the safety cost, starting from a $1$ to $100$ ratio with
respect to the asymptotic value and of the initial cost with respect to the
initial wealth.}
\label{fig-3}       
\end{figure}

\begin{figure}
\centering
\includegraphics[width=12cm]{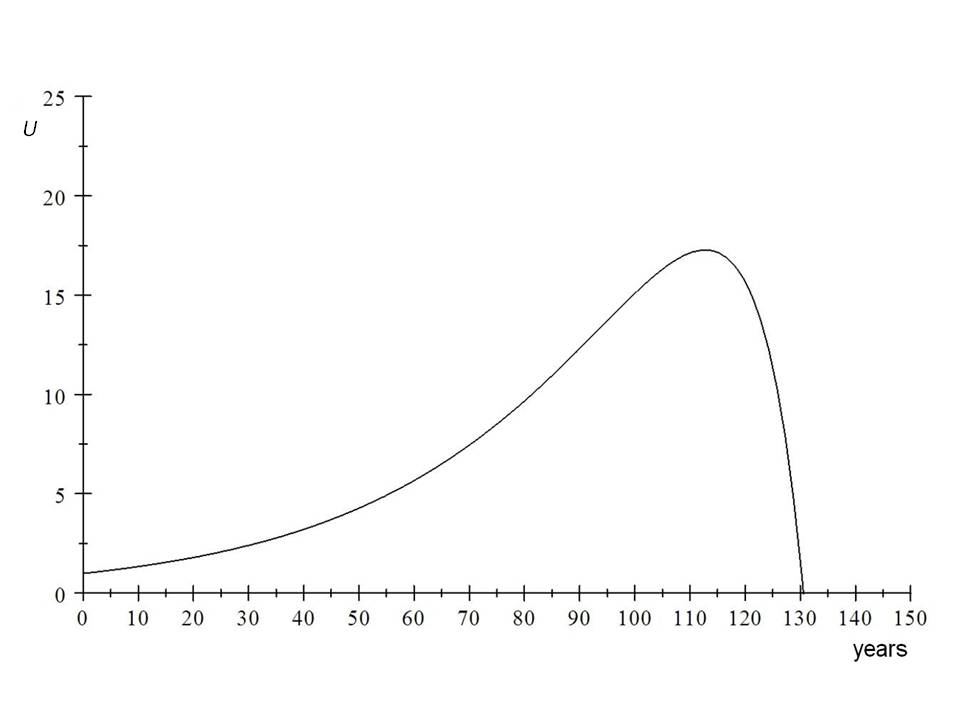}
\caption{Time evolution of the net
wealth given by the difference between the gross wealth produced by the
growing system and the cost of safety.}
\label{fig-4}       
\end{figure}

The most interesting portrayal is obtained when graphing the difference, $%
\mathcal{U}$, between wealth and safety cost, normalized to the initial
wealth; see figure (\ref{fig-4}).

Looking at the diagram a quotation from Leonardo da Vinci comes
spontaneously to my mind:

\begin{center}
\textit{Chi scalza il muro, quello gli cade addosso.\footnote{%
"If you dig up the wall, it will fall upon you". Free translation by the
author.}}
\end{center}

During about 115 years the net wealth or profit, $\mathcal{U}$, grows but in
the end it plunges catastrophically in more or less fifteen years. Of course
the result depends on the numerical values I have chosen. Keeping the
"wonderful" $3\%$ fixed, what is the change if I modify the ratio between
the initial and the asymptotic risk? Let's see.

\begin{figure}
\centering
\includegraphics[width=12cm]{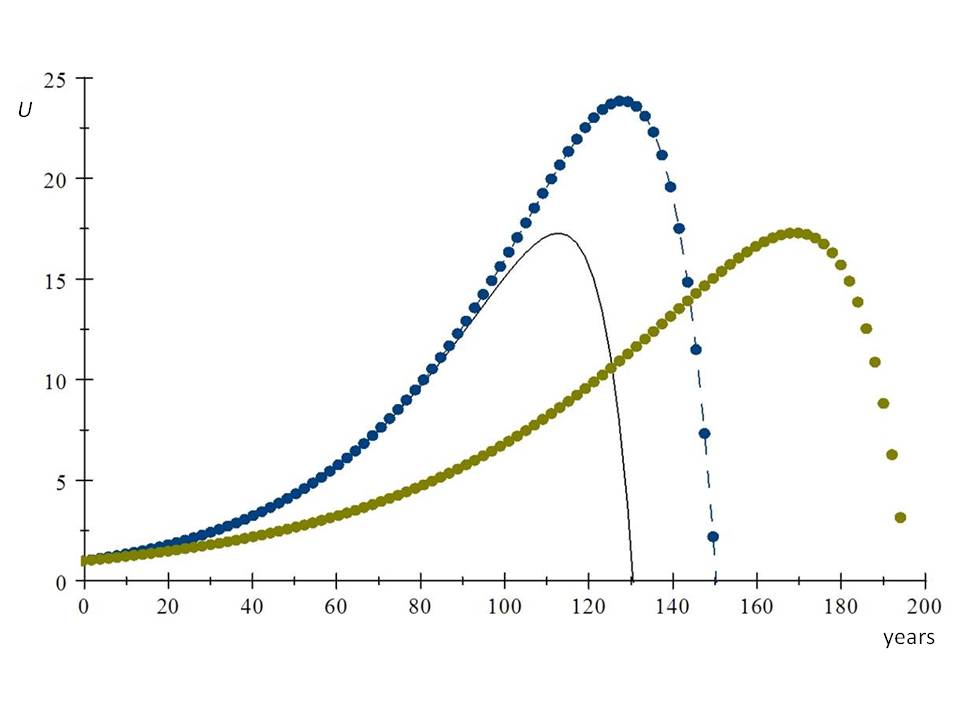}
\caption{Black curve: reference case.
Blue curve: $\protect\varepsilon _{0}/\protect\varepsilon _{\ast }=1000$.
Greenish line: $\protect\chi =0.02$.}
\label{fig-5}       
\end{figure}

Even assuming a $1000:1$ ratio between initial and asymptotic risk, the
trend remains the same and the "gain" in time is around twenty years. If
instead the annual growth rate is limited to $2\%$ the gained years are
about sixty.

This is just an exercise. The real world is far more complicated than that,
but ...

A trend like the one depicted in fig.~(\ref{fig-4}) is often found in the
history of human societies, from the Sumerian civilization to the Roman
empire, to many other cases. The example of the Sumerians is especially
interesting because, in their case, the collapse was due to the unforeseen
(and unforeseeable at the time) side effects of a great technical advance.
The development of a sophisticated irrigation system in a dry climate
brought, because of evaporation, to irreversibly salt the soils and led
their productivity to collapse. The productivity of the former Sumerian land
remains quite low even today, after four thousand years.

The typical feature is a blooming, almost triumphal, raise followed by a
much shorter and steeper descent:

\begin{center}
\textit{...the raise is gradual, the ruin is precipitous... }(Lucius Anneus
Seneca, Letters to Lucilius, 91-6\footnote{%
... incrementa lente exeunt, festinatur in damnum.}).
\end{center}

We could call it, as Ugo Bardi does, "Seneca effect". For completeness I
should mention that he also introduces a "Cassandra effect" that we would
like to avoid here.

Besides the historical examples are there circumscribed phenomena where
mechanisms are operating like the ones I described above? Let us think to
the role (and cost) of bureaucracy in our society. Let us think to the
development of e-mail networks and to the percent of our work time we must
dedicate to the handling of the correspondence: each communication is very
quick, the answer time is not. Let us think to informatization and to the
queues before the counters of public offices: the length of the queue does
not decrease, the number of counters does.

\subsection{Saturation and capacity of keeping processes under control}
\label{sec-43}

In our idealized network of exchange fluxes I have pretended to believe that
each single vertex can in principle be connected with any other vertex of
the net: this hypothesis is the base of (\ref{relazioni}). However, as we
have seen, both the exchange channels and the vertices get saturated, which
fact leads to transferring the growth to the total number of vertices.
Unavoidably then it becomes gradually impossible for a vertex, even in
principle, to directly communicate with \textit{all} other vertices. This
fact does not prevent the communication with the parts of the net out of
reach, but implies that the related flux be mediated by the vertices with
which the link is direct; the indirect flux must however be transmitted
through the vertices with which one is in direct touch. Every vertex, then,
must support its own exchange flux together with the flux in transit. The
latter then shortens the saturation times of any single vertex. In other
words we can say that a competition is established concerning the allocation
of time to control your flux against the flux in transit; otherwise stated,
the more are the vertices used for simple transit, the slower becomes the
distant exchange. Summing up, the myth of globalization must face the
problem of control: the growth of the network implies both a parceling of
the system in governable islands and a functional hierarchy of the vertices.

A complicated way to say that an upper limit exists to the physical size of
a single living being composed of cells: there have never existed
multicellular animals bigger than the blue whale (maybe equalled, but not
surpassed, by some terrestrial dinosaur of the Jurassic period). With a
growing size the system goes out of control by a single will.

So far I have exorcised a word I could and maybe I should have used:
entropy. It is a physical quantity introduced by thermodynamics, that can
easily be misused. Anyway, the popular version of its definition says that
it is something having to do with disorder in a system made of a very big
number of components; and this is true. Furthermore one of the
non-negotiable laws asserts that in an isolated complex system whatever
happens forces entropy, then disorder, to grow. The surface layer of our
planet is not an isolated system, but it exchanges with the external space
almost only radiation. If the number of physical processes taking place
annually on earth remains more or less stable, so does the entropy and this
is because it is possible to expel energy to space. However if the number of
thermodynamic transformations increases, then, in order to keep the entropy
stable, the surface temperature of the planet \textit{must }grow. The
phenomenon about which I am speaking now is \textit{not} the world famous
green-house effect, on which the discussion will be tomorrow: rather it is
an additional complication.

So far the most traditional meaning. It is however possible to define a
statistical entropy associated to complex systems irrespective of strictly
thermal variables. Order and disorder of the books on a shelf can be
evaluated in terms of entropy and it is true that, without "external"
interventions for re-ordering, the more books you draw and use the more
disorder grows. Restoring or maintaining order requires a dedicated work,
that can be assessed in terms of time (to be subtracted from the productive
time) then finally in terms of money. We come back, by another way, to the
Seneca effect.

\section{Treating the addict}
\label{sec-5}

I have touched upon a number of problems and effects, that could be analyzed
in more formal and technical terms. The relevant feature, however, is that
they are not debatable, even though there are many uncertainties in their
formal description and in their parametrization.

Well, no discussion nor, least of all, any account is taken of them in the
sites where decisions relevant for the community are taken; excepting, at
most, fortuitous actions on symptoms when acute effects appear.

A comparison that comes to my mind is with a doctoral team facing an addict
or an alcoholic in a drug withdrawal syndrome. The patient is sick, very
sick, because he cannot find the dose. Suffering is real and the danger of
irreversible damages is real too. Everybody knows that, in this situation,
the sheer block of the assumption of the dependence giving substance is not
viable; it is necessary to define a proper exit strategy that includes a
gradual elimination of the substance, accompanied by a controlled
administration of some less toxic substitute and by protective and healing
measures. In any case the final goal is to free the patient from his
dependence.

However in the case of the "growth" disease the medical team, up against
recurring withdrawal crises, has only one strategy: to strive for providing
the patient with the dose he misses and, since the dependence increases more
and more, to try and procure in turn increasing doses. The final result of
such a "therapy" is rather evident.

The fact that makes things harder is however that in our case the patient
and  the medical team essentially coincide.

Those who professionally deal or have the misfortune to care alcoholics or
drug addicts perfectly know one thing. That any exit strategy from a
dependence state by means of a withdrawal therapy requires an inescapable
preliminary condition: the patient must agree. And this is the mess.

\section{Reason or scope intelligence?}
\label{sec-6}

Human beings have a bizarre feature: their mind is equipped with analysis
tools of the reality which enable, in a reasonably reliable way, to forecast
the consequences of their behaviours even over medium/long times and over
distances  much bigger than the ones they can directly inspect by means of
their senses. Let us call this ability: \textit{reason}.

Human beings however, to some extent, share with the other animals an
apparently similar ability, which however does not bother about the future,
if not immediate, and about the space range, if not directly perceived by
the senses. This ability I could call "scope intelligence".

We know that chimpanzees are able to build simple tools, in order, for
instance, to facilitate the extraction of honey from honeycombs hosted
within holes in the logs of trees. Crows too have learnt, in order to smash
nut shells, to bring them flying at a given height then letting them fall
against a hard surface. Examples are abundant and describe the nature of
what I have called \textit{scope intelligence}.

Let us now imagine some chimpanzees who, keeping the purpose to pull out
honey from logs (or to procure everyday's food in similar situations) learn
to build not only simple wooden sticks to be introduced into the cavity then
licked, but drills, diggers, explosives and so on. This peculiar species of
"advanced" chimpanzees would rapidly reach the depletion of the food
resources of its environment, ending then in its own extinction.

Specific human civilizations have indeed concluded their life like that,
collapsing under the unforeseen effects of some progress introduced thanks
to the scope intelligence and in the helplessness of reason: I have already
mentioned the example of Sumerians and the salinization of their
agricultural land, source of their prosperity.

Today's situation, however, is still different, because reason is now able
to predict many effects of an expansion of economy that pretends to last
indefinitely; but this type of prediction is either ignored or rejected. We
are facing a real conflict between reason and scope intelligence, between
human and animal nature, and the match risks to be compromised by a fact
ironically expressed by Albert Einstein who happened to remark that

\begin{center}
\textit{the difference} \textit{between stupidity and genius is that genius
has its limits.}
\end{center}

Notwithstanding the difficulties, however, most of the international
scientific community believes that we can make the grade and that reason
will in the end have the upper hand on the scope intelligence.

Our condition has very well been described in what we could call a laic
parable told in the 19th century by  Robert Louis Stevenson: \textit{Strange
case of Dr Jekyll and Mr Hide}. Dr Jekyll is a reasonable and respectable
person, he is aware of what is wise to do; however the man who takes
decisions on what to do is Mr Hide. One might think, and very often some
think so, that it would suffice for Jekyll to get rid of Hide and everything
would be solved, but when one tries that way tragically realizes that Dr
Jekyll and Mr Hide are the same person...

This is a conference for Dr Jekyll, who assembled here in order to clarify
his mind on what to do. The point however is that afterwards it will be
necessary to bridle Mr Hide. We are going to tackle this problem again at
the end of our four days.

\end{document}